\documentclass[%
 reprint,
 amsmath,amssymb,
 prl,
 superscriptaddress,
]{revtex4-1}

\usepackage{natbib}
\usepackage{amsmath, amssymb}
\usepackage{fancyhdr, ifpdf}
\usepackage{bm}
\usepackage{cases}
\usepackage{color}

\usepackage{braket}
\usepackage{graphicx}
\usepackage{dcolumn}
\usepackage{bm}

\begin{document}

\title{Vacuum Rabi spectra of a single quantum emitter}


\author{Yasutomo Ota}
\email{ota@iis.u-tokyo.ac.jp}
\affiliation{Institute for Nano Quantum Information Electronics, The University of Tokyo, 4-6-1 Komaba, Meguro-ku, Tokyo 153-8505, Japan}%

\author{Ryuichi Ohta}
\affiliation{Institute of Industrial Science, The University of Tokyo, 4-6-1 Komaba, Meguro-ku, Tokyo 153-8505, Japan}%

\author{Naoto Kumagai}%
\affiliation{Institute for Nano Quantum Information Electronics, The University of Tokyo, 4-6-1 Komaba, Meguro-ku, Tokyo 153-8505, Japan}%

\author{Satoshi Iwamoto}%
\affiliation{Institute for Nano Quantum Information Electronics, The University of Tokyo, 4-6-1 Komaba, Meguro-ku, Tokyo 153-8505, Japan}%
\affiliation{Institute of Industrial Science, The University of Tokyo, 4-6-1 Komaba, Meguro-ku, Tokyo 153-8505, Japan}%

\author{Yasuhiko Arakawa}%
\affiliation{Institute for Nano Quantum Information Electronics, The University of Tokyo, 4-6-1 Komaba, Meguro-ku, Tokyo 153-8505, Japan}%
\affiliation{Institute of Industrial Science, The University of Tokyo, 4-6-1 Komaba, Meguro-ku, Tokyo 153-8505, Japan}%

\date{\today}

\begin{abstract}
We report the observation of the vacuum Rabi splitting of a single quantum emitter by measuring its direct spontaneous emission into free space. 
We used a semiconductor quantum dot inside a photonic crystal nanocavity, in conjunction with an appropriate cavity design and filtering with a polarizer and an aperture, enabling the extraction of the inherently-weak emitter's signal. 
The emitter's vacuum Rabi spectra exhibit clear differences to those measured by detecting the cavity photon leakage.
Moreover, we observed an asymmetric vacuum Rabi spectrum induced by interference between the emitter and cavity detection channels. 
Our observations lay the groundwork for accessing various cavity quantum electrodynamics phenomena that manifest themselves only in the emitter's direct spontaneous emission. 
\end{abstract}
\maketitle
Cavity quantum electrodynamics (QED) studies the interaction between cavity photons and quantum emitters, such as Rydberg~\cite{Raimond2001} and neutral atoms~\cite{Boca2004}, superconducting qubits~\cite{Wallraff2004}, nitrogen vacancy centers in diamond~\cite{Park2006a} and semiconductor quantum dots (QDs)~\cite{Yoshie2004,Reithmaier2004,Hennessy2007a,Englund2007,Srinivasan2007}.
The emitter and cavity photons interact not only with each other at a rate of $g$, but also independently with the free-space vacuum field, leading to irreversible radiation at rates of $\gamma$ (for emitter) and $\kappa$ (for cavity).
When $g \gtrsim \kappa + \gamma$ and the pure emitter dephasing is negligible, the strong coupling regime is achieved and vacuum Rabi splitting (VRS) can be observed in the spectral domain. 

\begin{figure} 
\centering 
\includegraphics[width=\linewidth]{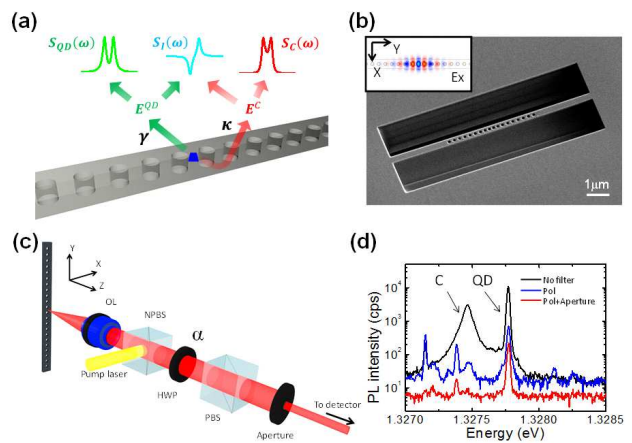}
\caption{(Color online)  
(a) Schematic illustration of the photon leakage from our cavity QED system. 
The direct QD emission into free space excites $E^{QD}$ through the $\gamma$ channel that results in the QD spectrum, $S_{QD}(\omega)$.
The cavity also spontaneously radiates and excites $E^{C}$ from the $\kappa$ channel, resulting in the cavity spectrum, $S_{C}(\omega)$.
Interference between the two leakage channels also occurs, resulting in $S_{I}(\omega)$.
Here, $\kappa$ and $\gamma$ include the leakage into both free space and the adjacent waveguide modes, and respectively reflect the total leakage rate of the cavity and emitter. 
Some part of the leaked photons direct toward the detector, conceptually placed above the nanobeam cavity.
(b) Scanning electron beam micrograph of our device. The inset shows the cavity field distribution overlaid on the cavity design.
(c) Schematic of the experimental setup. The sample is addressed by a microscope objective lens (OL) and pumped by a continuous wave laser reflected by the non-polarized beam splitter (NPBS).
The collected light passes through a halfwave plate (HWP), polarizing beam splitter (PBS) and an aperture, and then sent to a spectrometer.
The HWP angle, $\alpha$, is defined as the angle between the HWP fast axis and the x axis (which is parallel to the horizontal axis of our optical table). 
(d) $\mu$PL spectra showing the effect of the polarizer and aperture. 
Without any filtering, bright cavity peak (C), together with a sharp QD peak (QD), can be seen (black curve). 
The rotation of the HWP to $\alpha = 50.5^\circ$ strongly suppresses the cavity contribution in the spectrum (blue curve). 
Further cavity mode suppression can be obtained by the insertion of the aperture (red curve).
}
\label{fig:f1}
\end{figure}

In general, the two radiation channels exhibit different spectra, which, in principle, can be separately measured as the emitter ($S_{QD}(\omega)$) and cavity ($S_{C}(\omega)$) VRS spectra, as illustrated in Fig. \ref{fig:f1}(a). 
Moreover, their interference ($S_{I}(\omega)$) should be expected to affect the measured spectral shape. 
However, to date VRS spectra have been measured predominantly by cavity transmission, reflection and emission spectroscopy~\cite{Boca2004,Wallraff2004,Yoshie2004,Reithmaier2004,Park2006a,Hennessy2007a}, since it is brighter and easier to access for most cavity QED systems. 
This is particularly the case for QD-based cavity QED systems, since most of these possess high cavity leak rates, such that $\kappa \gg \gamma$.

In atomic cavity QED systems, some spectroscopic studies on the emitter channel have been performed~\cite{Childs1996} but none of these were performed in the single emitter strong coupling regime. 
In contrast, theoreticians frequently discuss the emitter spectra~\cite{Carmichael1989,Loffler1997,Cui2006,Laussy2008,Yamaguchi2008,Yamaguchi2008a,Auffeves2008,Laussy2011}, the importance of which has been discussed pertaining to the study of several intriguing phenomena, including the quantum-classical crossover in single atom lasers~\cite{Loffler1997,DelValle2010}, quantum phase transitions in the Jaynes-Cummings Hubbard model~\cite{Knap2011}, cavity induced transparency~\cite{Alsing1992,Auffeves2008} in low $\kappa$ systems and quantum delayed-choice experiments~\cite{Stassi2012}. 
Importantly, most of these phenomena manifest themselves only in the emitter spectra. 
The significance of the interference between the two leakage channels (that has a similarity with Fano interference) has also been addressed in the literature~\cite{Yamaguchi2008a,Barclay2009,Madsen2013}.

In this study, we demonstrate an experimental measurement of VRS by detecting direct spontaneous emission from an emitter into free space, 
using a single InAs/GaAs QD strongly coupled to a photonic crystal nanobeam cavity. 
We show that the simple combination of an appropriate cavity design and filtering with both a polarizer and aperture enables the extraction of the emitter channel contribution from the free space radiation field, which is otherwise dominated by the cavity leakage. 
This technique allows us to isolate and measure the emitter spectra, $S_{QD}(\omega)$, and to compare it with $S_{C}(\omega)$, as well as to study the interference part, $S_{I}(\omega)$, that provides some asymmetry in the VRS spectra.

We fabricated the photonic crystal nanobeam cavity onto a low density quantum dot wafer grown by molecular beam epitaxy, using a standard combination of electron beam lithography, dry and wet etching. 
The details of the sample fabrication process can be found in our previous publication~\cite{Ohta2011}.
Figure~\ref{fig:f1}(b) shows a scanning electron microscope image of a fabricated device. 
The cavity is formed at the center of the air-bridge photonic crystal nanobeam by modulating the air hole patterning period~\cite{Ohta2011}, 
and is designed to have a moderate $Q$ factor of 49,000 with suppressed leakage to the vertical direction ($\sim 2 \%$ of the total radiation).
This suppression is achieved by tuning the number of reflecting air holes~\cite{Enderlin2012} and guiding dominant cavity leakage into the side waveguide (see Supplementary Section II~\cite{Sup}). 
The cavity field distribution overlaid with the cavity design is shown in the inset. 
For optical measurements, the sample was placed inside a temperature controlled optical cryostat and was kept at 3.1 K throughout the measurement. 
We use a continuous wave Ti:Sapphire laser oscillating at 860 nm to pump the sample with a fixed power of 7.3 $\mu$W (except when measuring the spectra shown in Fig.~\ref{fig:f3}(d)).
The pump creates excitons in the QD that radiate as a dipole and finally decay after exciting the electromagnetic field in either the emitter channel ($E_{QD}$) or cavity channel ($E_{C}$), as illustrated in Fig. \ref{fig:f1}(a). 
Note that the pump power used is well below the saturation pump power of the QD under the resonance with the cavity mode.

We used a micro-photoluminescence ($\mu$-PL) technique to address the individual sample, with some filters in the detection path, as shown in Fig. \ref{fig:f1}(c). 
The pump laser light was focused by an objective lens with a numerical aperture of 0.65, which is also used for collecting radiation from the sample.
Spectra were obtained by a grating spectrometer equipped with a CCD detector placed after the filters (spectral resolution = 13.5 $\mu$eV). 
A combination of a half wave plate (HWP) at a angle, $\alpha$, of 50.5$^\circ$ and a polarizing beam splitter (PBS) enables the rejection of the major cavity farfield, which is polarized roughly parallel to the x-axis. 
The aperture, which has a diameter of 0.75 mm and reduces the detection numerical aperture roughly to 0.1, is used to suppress the minor cavity farfield which is polarized orthogonally to the major one (see Supplementary Section III~\cite{Sup}).
The effect of these filters on the measured spectrum of a detuned QD-cavity system is shown in Fig.~\ref{fig:f1}(d). 
Without these filters, strong cavity mode emission is seen (black curve) but it can be largely reduced by the polarization filtering (blue curve). 
Further reduction is obtained by the aperture (red curve), leading to the near-perfect suppression of the cavity emission that is required to successfully extract $S_{QD}(\omega)$ near the emitter-cavity resonance (which would  otherwise be flooded out by $S_{C}(\omega)$).
In the following measurements, we rotate the HWP to compare the two spectrum components, $S_{QD}(\omega)$ and $S_{C}(\omega)$, but keep the aperture inserted. 

\begin{figure}
\centering
\includegraphics[width=\linewidth]{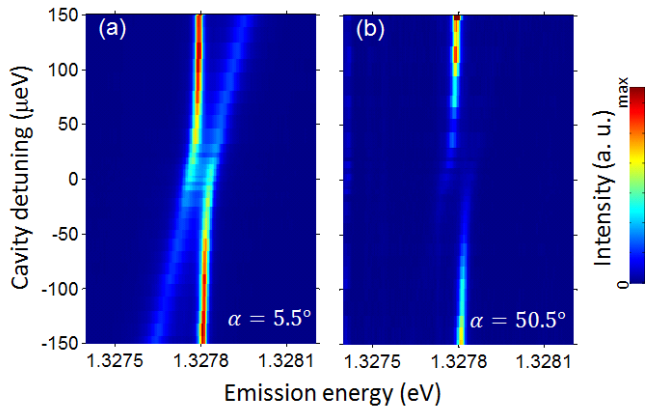}
\caption{(Color online)  
(a) Cavity channel spectra under various detunings. $\alpha$ is set to 5.5$^\circ$. The spectra contain weak contributions from the emitter channel leakage (see text). 
(b) Emitter channel spectra under various detunings. $\alpha$ is set to 50.5$^\circ$. 
}
\label{fig:f2}
\end{figure}

\begin{figure}
\centering
\includegraphics[width=\linewidth]{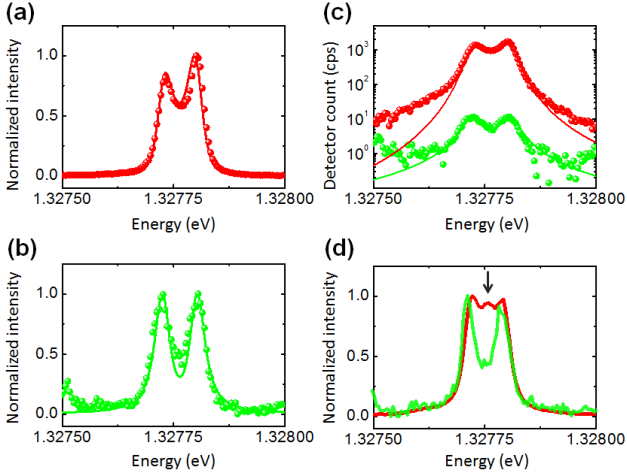}
\caption{(Color online) 
(a) Normalized spectra at the resonance measured when mostly emphasizing the cavity leakage contribution ($\alpha$ = 5.5$^\circ$), and (b) when accepting only the emitter channel contribution ($\alpha$ = 50.5$^\circ$). 
(c) Unnormalized spectra corresponding to (a) and (b). 
For these three panels, the balls are experimental data and the solid lines are of numerical simulations, details of which are presented in Supplementary Section V~\cite{Sup}.
(d) Normalized VRS spectra obtained under a three times stronger pumping power than in (a)-(c).
Red (green) balls are of the mainly-cavity (emitter) spectrum. 
The black arrow indicates the emergence of the additional peak for the cavity leakage channel. 
Slight asymmetry in the QD spectra is due to unintentional detuning induced by the increase of the pump power.
}
\label{fig:f3}
\end{figure}

First, we measured the effect of emitter-cavity detuning on the emission spectra at two different HWP angles, as plotted in Fig.~\ref{fig:f2}, by tuning the cavity resonant frequency using a Xe gas adsorption technique. 
When setting $\alpha$ = 5.5$^\circ$, we obtain spectra that mostly emphasizes the cavity emission contribution, $S_{C}(\omega)$. 
This situation is very similar to almost all previous $\mu$-PL experiments based on QD-based cavity QED systems. 
In Fig.~\ref{fig:f2}(a), the cavity mode-like emission, which forms a diagonal line in the colorplot, is visible even under detuned conditions due to off-resonant mode coupling that is prominent in many QD-based systems~\cite{Hennessy2007a}. 
An anti-crossing of the two emission peaks is clearly observed as the cavity is tuned into resonance with the QD, demonstrating that the system is in the strong coupling regime. 
In contrast, we observed a largely different behavior when measuring $S_{QD}(\omega)$ by setting $\alpha$ = 50.5$^\circ$ (see Fig.~\ref{fig:f2}(b)). 
The diagonal cavity-like emission line is not visible under detuned conditions and the total emission intensity significantly decreases as the system is tuned into resonance.
This is because $\kappa \gg \gamma$ in our system and the system energy decays dominantly through the cavity channel when near the resonance condition. 
Nevertheless, we still observe a clear anti-crossing in this detection geometry. 
A detailed comparison of the two set of spectra in terms of intensities, linewidths and peak positions can be found in Supplementary Section IV~\cite{Sup}.

Now, we discuss the VRS spectra at the resonance. 
Figure \ref{fig:f3}(a) shows a normalized resonance spectrum when $\alpha$ = 5.5$^\circ$. 
This spectrum is dominated by the cavity channel, and clearly shows a vacuum Rabi doublet with a splitting of 64 $\mu$eV. 
The emission spectra exhibits some asymmetry in that the higher energy emission peak is more intense.
The origin of this asymmetry is interference between the cavity leakage field and a minor contribution of the emitter field, namely due to the term, $S_{I}(\omega) = Re[E^*_{QD}E_C]$. 
This contribution from the emitter field arises because our QD is elliptically polarized and its direct spontaneous emission ($E_{QD}$) can not be rejected by the polarization filtering, resulting in the fact that we measure $S_{C}(\omega) + S_{I}(\omega)$  (see Supplementary Section I and V~\cite{Sup}).
However, by rotating $\alpha$ to 50.5$^\circ$, we can reject the contribution from the cavity leakage and can solely measure $S_{QD}(\omega)$, as plotted in Fig. \ref{fig:f3}(b).
The doublet now becomes symmetric and exhibits a deeper central dip than in the mainly-cavity spectrum. 
In addition, the VRS now shows a slightly wider splitting of 75 $\mu$eV with a wider spectral distribution in the exterior of the doublet.  
These features are characteristic for VRS spectra of the emitter channel under emitter-driven conditions~\cite{Cui2006,Laussy2011}.
We note that, since the emitter and cavity photon have a complementarity in this linear strong coupling regime, we should expect that the cavity VRS spectrum under cavity-driven conditions resembles the emitter VRS spectrum under emitter-driven conditions (see Supplementary Section VI~\cite{Sup}).

In these plots, the solid lines show the fitting of the data to our theoretical model, which carefully considers the detection geometry (see Supplementary Section V~\cite{Sup}).
In the model, we explicitly consider the pure dephasing of the QD for a better reproduction of the experimental results.
Most of the parameters needed for the calculation can be determined experimentally, such as $|g|$ = 41 $\mu$eV, $\kappa$ = 66 $\mu$eV ($Q \sim$ 20,000) and $\gamma$ =  0.28 $\mu$eV, 
leaving just three free parameters (emitter pure dephasing rate, cavity-free space coupling phase and degree of $E_{QD}$-$E_C$ field overlap). 
The experimentally obtained parameters accurately reproduce the size of the vacuum Rabi splitting measured through the two detection channels, using formulas derived by C. Cui and Raymer~\cite{Cui2006}, namely 
$2\sqrt{|g|^2-(\kappa ^2 +\gamma ^2)/8}$ (= 67 $\mu$eV) for the cavity channel and $2\sqrt{(|g|^4+2|g|^2\kappa(\kappa+\gamma)/4)^{1/2}-\kappa^2 /4}$ (= 75 $\mu$eV) for the emitter channel.
By setting these free parameters to a set of reasonable values, such as a pure dephasing rate of 3 $\mu$eV, we are able to reproduce the experimental results (including those discussed later).
Note that, in the model, we fixed the incoherent pumping rate of the QD to be 0.065 $\mu$eV in order to realize the week driving condition as in the actual experiments.
 
Figure \ref{fig:f3}(c) shows a logarithmic plot of unnormalized resonance spectra for the two detection channels.
The peak intensity for the emitter side is roughly 100 times weaker than the other. 
The simulation clearly reveals the strong asymmetry in the outer sides of the doublet of the mainly-cavity spectrum (red). 
Experimentally, however, the relatively strong background emission hindered observation of this, and the asymmetry is shown by the relative intensities of the doublet peaks.

In Fig \ref{fig:f3}(d), we show a spectrum taken under a roughly three times higher pumping power than that used for the rest of experiments.
Under this strong pumping condition, an additional peak between the VRS appears in the mainly-cavity spectrum (red curve), as indicated by the black arrow.
This third peak is often experimentally observed in QD-based cavity QED systems and is believed to arise from the QD's spectral blinking and also off-resonant cavity feeding~\cite{Hennessy2007a}.
On the other hand, the emitter side (green curve) still exhibits the emission doublet.
The contrast between the two spectra further supports the above scenario for the explanation of the origin of the triplet.
This result showcases the advantages of our measurement technique for discussing pure VRS spectra without being bothered by the third peak.

\begin{figure}
\centering
\includegraphics[width=\linewidth]{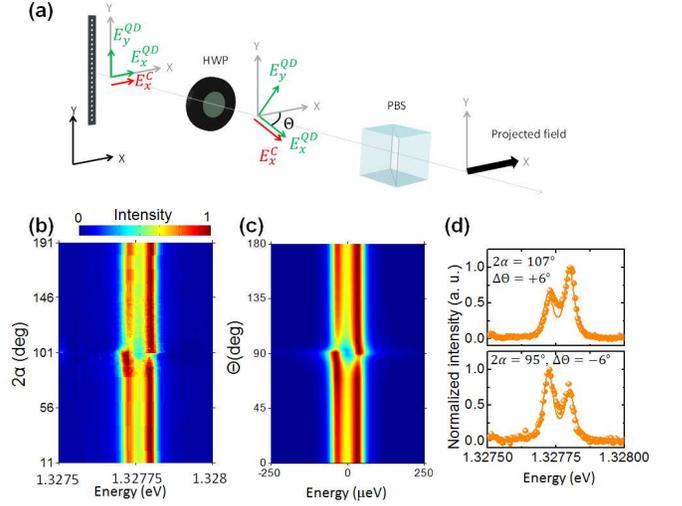}
\caption{
(Color online) 
(a) Schematic of the interference experiment. The radiation field from the emitter ($E_x^{QD}$, $E_y^{QD}$) and cavity ($E_x^C$) pass through the HWP and are projected onto the PBS with a angle, $\Theta$.
The PBS only allows the transmission of x-polarized light and the resulted field becomes a mixture of the emitter and cavity fields and is expressed as $E = E_x^C \cos(\Theta ) + E_x^{QD} \cos(\Theta ) + E_y^{QD} \sin(\Theta ) $.
(b) Measured VRS spectra for different HWP angles, $\alpha$. The spectra are mostly modified around $2\alpha \sim  101^\circ$, where the effect of the channel interference is most prominent. 
Each spectra for different $\alpha$ are normalized to its peak. 
(c) Calculated VRS spectra corresponding to (b). The spectra are calculated for different $\Theta$ and are normalized as in (b). 
(d) Two asymmetric VRS spectra highlighting the effect of channel interference. 
$2\alpha$ is set to 95$^\circ$ ($\Delta\Theta = -6^\circ$ ,lower panel) and 107$^\circ$ ($\Delta\Theta = +6^\circ$ ,upper panel). 
Solid lines show of the numerical simulation.
}
\label{fig:f4}
\end{figure}

Next, we apply our technique to study the influence of the channel interference.
We examine how the interference modifies the spectra simply by rotating the HWP and mixing the two channel contributions in a controlled manner. 
A schematic illustration in Fig.~\ref{fig:f4}(a) explains the details of the experiment. 
Our elliptically polarized QD couples to free space by exciting the $E_x^{QD}$ and $E_y^{QD}$ fields, while the cavity is assumed to contribute only by its coupling to the $E_x^C$ field.
By rotating the HWP, we can control the field projection angle, $\Theta$, defined with respect to the x axis. 
Only x polarized light is transmitted by the PBS, such that the projected field after the PBS is a mixture of the emitter and cavity channel contributions and is expressed as $E = E_x^C \cos\Theta + E_x^{QD} \cos\Theta + E_y^{QD} \sin\Theta $.
The measured spectra is proportional to $E^*E$, which, thereby, contains $\Theta$-dependent interference terms $\propto Re[E_x^{C*} E_x^{QD}], Re[E_x^{C*} E_y^{QD}]$.
When $\Theta = 90^{\circ} + \Delta \Theta$ with $\Delta \Theta \sim 0$ and using the fact $|E_x^{C}| \gg |E_y^{QD}| \sim |E_x^{QD}| $, the detected field can be approximated as $|E|^2 \sim |E_y^{QD}|^2 - 2\Delta \Theta Re[E_y^{QD*} E_x^C]$, 
and therefore we will observe both the pure emitter spectrum and a significant contribution from the interference term, $Re[E_y^{QD*} E_x^C]$.
This term is originally zero, since it arises from the orthogonally-polarized fields, but becomes finite due to the field projection onto the polarizer. 

Figure~\ref{fig:f4}(b) shows a color plot of normalized PL spectra at resonance taken whilst varying $\alpha$ from 5.5$^\circ$ to 95.5$^\circ$, which corresponds to
a change of $\Theta$ from 0$^\circ$ to 180$^\circ$.
A complicated change in the spectra can be seen, especially around $2\alpha \sim 101^\circ$ ($\Theta \sim  90^\circ$). 
Numerical calculations for different $\Theta$s were also performed, and the results plotted in Fig.~\ref{fig:f4}(c). 
The agreement between the two set of spectra is remarkable, and the interference effect is highlighted when $2\alpha$ is set to 95$^\circ$ ($\Delta \Theta =  -6^\circ$) and 107$^\circ$ ($\Delta \Theta =  +6^\circ$), as shown in Fig. \ref{fig:f4}(d).
Significantly asymmetric VRS spectra are clearly observed, and the asymmetry flips between $\Delta \Theta = \pm 6^\circ$. 
This is readily explained by the fact that the interference spectrum is very asymmetric (as illustrated in Fig.~\ref{fig:f1}(a)) and only the sign of its contribution flips when the sign of  $\Delta \Theta$ changes.
This observation clearly indicates the importance of the channel interference on VRS spectra. 
We note that the observed spectra contains information of the phase relationship between the cavity, emitter and the free space.
By a comparison to our theoretical model, we estimate the relative phase between the cavity and free space field to be roughly $\sim$ 0$^\circ$ (see Supplementary Section V~\cite{Sup}).
This capability to know the cavity/free space relative phase will be valuable for deeper understanding of open quantum system theories, 
as it is often hard to theoretically determine its value due to the nontrivial form of the coupling Hamiltonian~\cite{Viviescas2003}.

In summary, we have measured the VRS spectra of a single quantum emitter and observed spectral modification due to leakage channel interference. 
We show that a simple combination of an appropriate cavity design, a polarizer and an aperture is useful for extracting the emitter's direct spontaneous emission into free space, as well as for controlling the interference between the two detection channels.
Our demonstration will provide a means to access various intriguing quantum optics phenomena that manifest themselves in only the emitter spectrum, such as the incoherently-pumped Mollow triplet, which is predicted to occur when single emitter cavity QED systems start lasing~\cite{Loffler1997,DelValle2010}. 
In particular, such phenomena in nonlinear strong coupling regimes of cavity QED are of interest for further experimental studies, since they could be hard to reproduce using many-atom cavity QED systems~\cite{Childs1996,Zhu1990}.
We hope that our findings also stimulate discussion on novel cavity QED experiments and theories that use the emitter direct spontaneous emission, which, for example, may act as an novel feedback channel in quantum control.


\begin{acknowledgments}
The authors thank M. Holmes, S. Kako, K. Kamide and M. Yamaguchi for fruitful discussions.
This work was supported by Project for Developing Innovation Systems of the Ministry of Education, Culture, Sports, Science and Technology (MEXT), Japan, and by the Japan Society for the Promotion of Science (JSPS) through its Funding Program
for world-leading Innovation R$\&$D on Science and Technology (FIRST Program).
\end{acknowledgments}

\renewcommand{\thefigure}{S\arabic{figure}}
\renewcommand{\theequation}{S\arabic{equation}}
\def\vector#1{\mbox{\boldmath $#1$}}
\newcommand{\vectorII}[1]{\scriptsize{\vector{#1}}}

\section*{Supplementary Information}

\subsection*{I. Quantum dot characterization}

\begin{figure}
\centering
\includegraphics[width=\linewidth]{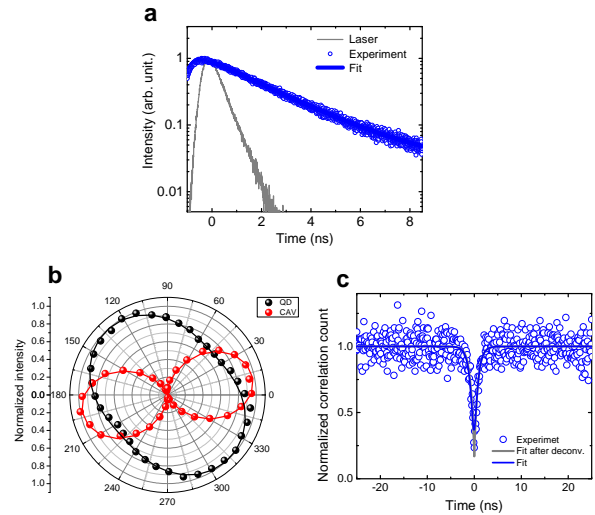}
\caption{{\bf Quantum dot emission properties.} 
{\bf a}, Time resolved PL for the QD emission line. The emission life time was evaluated as 2.35 ns (0.28 $\mu$eV).
{\bf b}, Normalized PL intensities plotted as a function of 2$\times$ the HWP angle. Balls (lines) are experimental (fitting) data.
Black (red) curves are of QD (cavity). 
{\bf c}, Measured intensity correlation function for the QD emission line. Circles are experimental data.
Blue and gray lines are respectively of a fitting curve and that after deconvolution with the detector response function (time resolution = 0.7 ns).
}
\label{fig:qd}
\end{figure}

Here, we characterize the QD emission used in the experiments under a far-detuned condition, where the cavity is red shifted by 5.7 meV from the QD emission line.
For the experiments in this subsection, we removed the aperture. 
First, we performed time resolved PL measurement on the QD using a pulse laser source (pulse width $\sim$ 1 ps) centered at 860 nm and a single photon counter.
The result is plotted in Fig.~\ref{fig:qd}(a). 
The emission life time was evaluated to be 2.35 ns (0.28 $\mu$eV) by a single exponential fit after convolving with the detector response function (time resolution = 0.45 ns). 
Since the QD is far detuned from the cavity, this value can be considered as the spontaneous emission life time without the Purcell effect, which is roughly
twice slower than the average of those for QDs in unprocessed area (1.2 ns). 
This weak photonic bandgap effect compared to two dimensional photonic crystals is one of advantages for the usage of the one dimensional photonic crystal nanobeam cavity,
since it has a larger spontaneous emission rate into free space, which makes the emitter channel measurements brighter and easier. 
Next, we discuss the polarization properties of the QD dipole. 
We consider that the emission of the QD arises from the positively-charged exciton, which is the dominant exciton species created in our QD wafer~\cite{Kumagai2010}.
This speculation is also supported by the fact that the QD emission line shows a linear pump power dependence and does not show fine structure splitting.
According to S. Ohno et al.~\cite{Ohno2011}, the QD dipole, $\vector{d_{a}}$, of the positively-charged exciton with a weak heavy hole-light hole mixing emits elliptically-polarized light and has a form as follows, 
\begin{equation}
\vector{d_{a}} = |\vector{d_{a}}| ( \vector{\hat{x}} \cos \theta_{a} + e^{\pm i \phi_{QD}} \vector{\hat{y}} \sin \theta_{a}  ),
\label{eq:dipole}
\end{equation}
where $\vector{\hat{x}}$ and $\vector{\hat{y}}$ are unit vectors respectively parallel to x and y axes.
The two signs of $\phi_{QD}$ correspond to two possible states in the positively-charged exciton.
Figure~\ref{fig:qd}(b) shows measured PL intensities of the QD emission plotted as a function of 2$\times$ the HWP angle.
The data is taken under the same continuous wave pumping condition as the experiments in the main text. 
The plot also contains a measured curve for the cavity mode emission.
The curve for QD are fitted with a function $|\cos\theta_a\cos 2\alpha + e^{\pm i \phi_{QD}}\sin\theta_a\sin 2\alpha |^2$, where $\alpha$ is the HWP angle.
Through the fit, we obtain $\theta_a$ = 42.6$^\circ$ and $\phi_{QD}$ = 80.8$^\circ$. Here, we set the x axis parallel to the cavity main axis (2$\alpha$ = 21.9$^\circ$).
Finally, we measured the intensity autocorrelation of the QD emission under a different detuning condition (the same as that used in Fig.1(d)). 
The result is shown in Fig.~\ref{fig:qd}(c). 
The observed anti-bunching with the time origin value of 0.17 demonstrates that our cavity QED system is working in the quantum regime.

\begin{figure*}[!htbp]
\centering
\includegraphics[width=15cm]{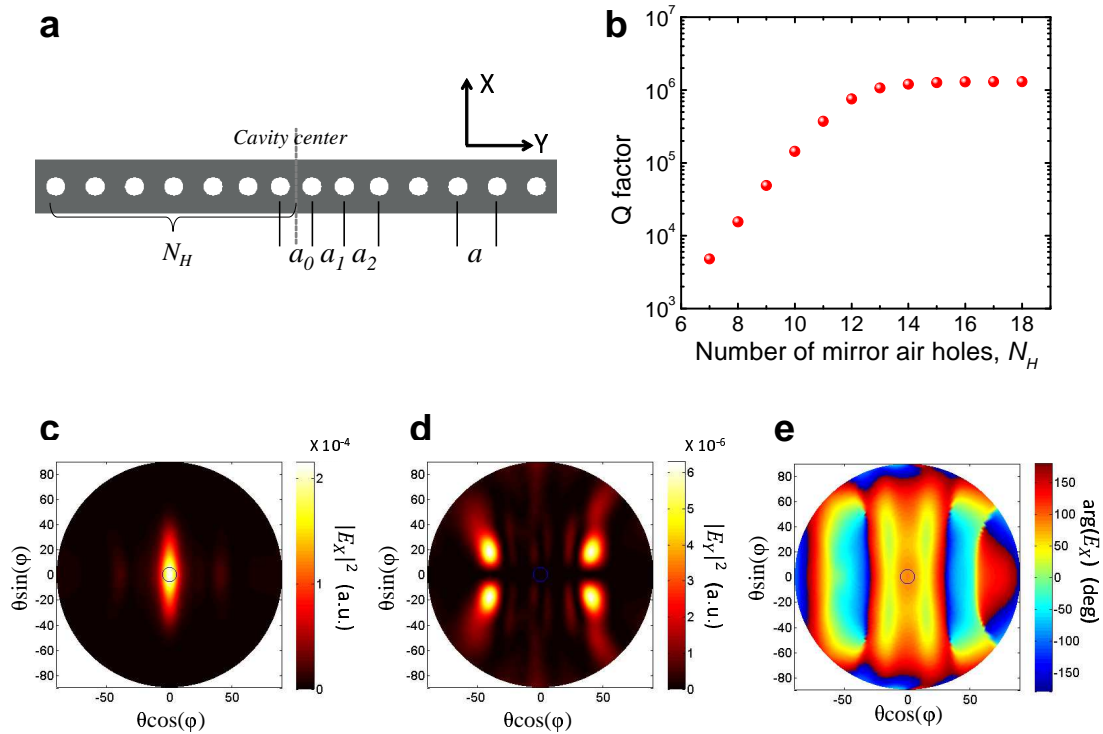}
\caption{{\bf Cavity mode characterization.} 
{\bf a}, Cavity design. 
The cavity center is indicated by the dashed line. 
Positions of the nearest neighbor airholes from the cavity center were tuned as follows: $a_0$=$0.8a$, $a_1$=$0.822a$ and $a_2$=$0.889a$. 
$N_H$ is number of mirror air holes. 
{\bf b}, Calculated cavity $Q$ factors for different $N_H$s.
{\bf c}, Intensity distribution of the $E_x$ farfield in the upper hemisphere of radiation space. 
$\theta$ is the polar angle and $\phi$ is the azimuthal angle in the spherical coordinate.
{\bf d}, The same in {\bf c} but for the $E_y$ farfield.
{\bf e}, Phase distribution of the $E_x$ farfield, in other words, a plot of  $\arg (E_x)$.
}
\label{fig:FF}
\end{figure*}

\subsection*{II. Cavity mode design and characterization}

We used a cavity design based on our previous work~\cite{Ohta2011,Enderlin2012}.
The structure is schematically shown in Fig.~\ref{fig:FF}(a), and has a base lattice constant ($a$) of 250 nm, slab thickness ($d$) of 130 nm, slab width ($w$) of 340 nm and an air hole radius ($r$) of 62.5 nm.
The refractive index of the GaAs ($n$) is assumed to be 3.4. 
The positions of the nearest neighbor holes from the cavity center are shifted as illustrated in the figure. 
We simulated the near- and far-field of the designed cavity by a finite difference time domain algorithm provided by Rsoft corporation. 
In order to suppress the light leakage into the vertical direction (perpendicular to the xy plane), we reduce the number of mirror air holes and guide the leaky light into the nanobeam waveguide.
Figure~\ref{fig:FF}(b) shows the simulated cavity $Q$-factors as a function of the number of mirror holes. 
By setting the hole number to 9, we have a high enough $Q$ factor of 49,000 while suppressing the vertical-top radiation to only 2 $\%$ of the total radiation. 
In this case, $\sim$ 94 $\%$ of light is channeled into the lateral waveguide.
The cavity also supports a small mode volume of only 0.29 $(\lambda/n)^3$.
We then calculated the farfield emission pattern by a near-to-far field conversion.
The results are plotted in Figs.~\ref{fig:FF}(c) and (d).
The $E_x$ farfield is concentrated in the center, while the $E_y$ farfield is much weaker and found at larger angles.
Thus, the $E_y$ field will be rejected by the insertion of the aperture.
We also calculated the phase of the $E_x$ farfield, as can be found in Fig.~\ref{fig:FF}(e).
Within the numerical aperture of 0.1, the phase of the field is about - 85$^\circ$.
This phase corresponds to that which the farfield acquires when escaping from the cavity.

\subsection*{III. Effect of aperture}

\begin{figure}
\centering
\includegraphics[width=\linewidth]{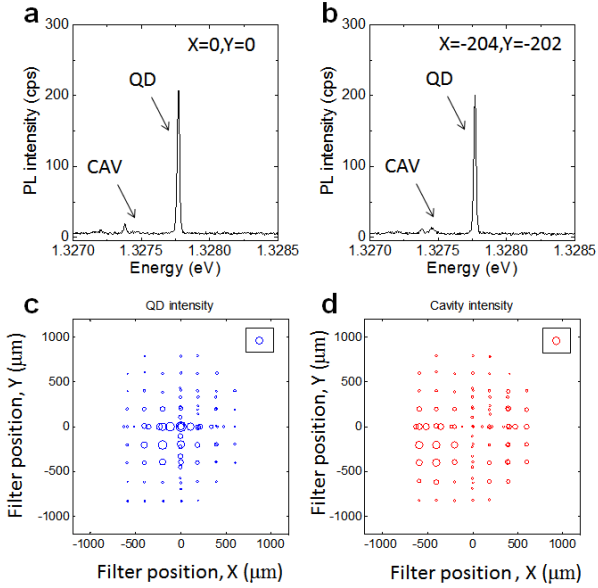}
\caption{{\bf Effect of aperture.} 
{\bf a}, Emission spectrum taken at the aperture position center, $X$=0 $\mu$m, $Y$=0 $\mu$m. 
The aperture center is set to the position used for the experiments in the main text, where the maximum ratio of the QD emission to the cavity emission is achieved. 
{\bf b}, Emission spectrum taken at an aperture position, $X$=-204 $\mu$m, $Y$=-202 $\mu$m. Weak cavity emission is visible.
{\bf c}, Integrated QD emission intensities plotted as a function of aperture positions. 
Intensities are expressed by the diameter of the circle. The inset corresponds to 1000 cps. 
XY axes here are parallel to the xy axes in Fig.~1(c). 
Note that the aperture has a round shape and the diameter is 750 $\mu$m. 
{\bf d}, The same in {\bf c} but for the cavity emission. The inset corresponds to 100 cps.
}
\label{fig:aperture}
\end{figure}

Here, we discuss how the aperture in the detection path affects the suppression of the cavity leakage contribution reaching to the detector.
We measured spectra at various aperture positions controlled by a motorized stage.
Continuous wave pumping at 860 nm with a power of 3.6 $\mu$W is used here and the HWP angle is fixed to 50.5$^\circ$ that maximally suppresses the $E_x$ cavity field intrusion, 
while the $E_y$ cavity field passes through.
Representative measured spectra are shown in Fig.~\ref{fig:aperture}(a) and (b). 
At the filter center (Fig.~\ref{fig:aperture}(a)), strong QD emission is observable while the cavity peak is fully suppressed.
Meanwhile, the cavity emission becomes visible when the aperture is deviated from the center by a few hundred $\mu$ meters (Fig.~\ref{fig:aperture}(b)).
The small peak next to the cavity peak is from another irrelevant QD.
Next, we compared the integrated intensities of the QD emission and that of the cavity mode at various aperture positions, as in Figs.~\ref{fig:aperture}(c) and (d).
Around the position center, QD emission becomes stronger while the cavity emission is largely suppressed. 
The cavity field intensity distribution that is weak around the center resembles that of the minor ($E_y$) cavity farfield (See Fig.~\ref{fig:FF}(d)). 
This suggests that the aperture mainly helps in suppressing the $E_y$ cavity farfield that can not be rejected by the polarization filtering.

\subsection*{IV. Detuning dependence of VRS spectra}

\begin{figure}
\centering
\includegraphics[width=\linewidth]{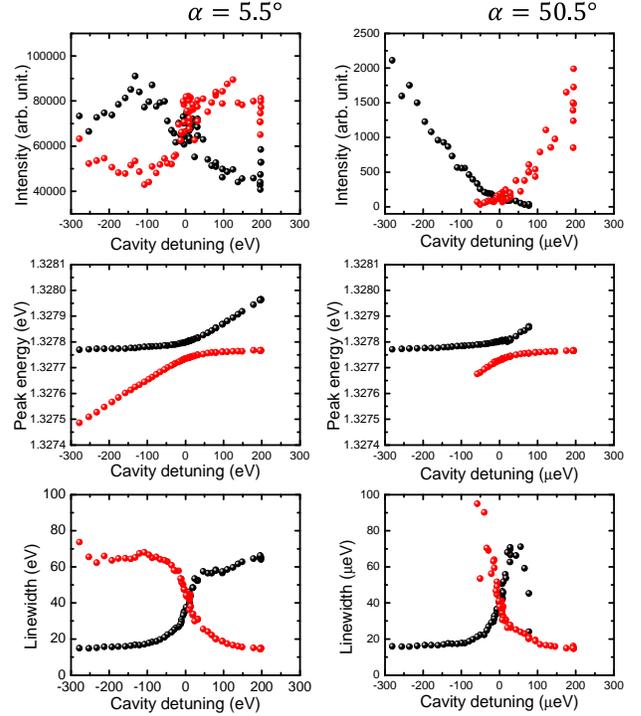}
\caption{{\bf Fitting to VRS spectra presented in Fig. 2.} 
Left panels are data for the mainly-cavity emission taken with a HWP angle of 5.5$^\circ$: 
(top) integrated intensities of the respective peaks, (middle) peak energies, and (bottom) linewidths.
Right panels are for the emitter spectra taken with a HWP angle of 50.5$^\circ$.
Red (black) balls are of lower (upper) polariton branch.}
\label{fig:fit_f2}
\end{figure}

Here, we show fitting results on the spectra displayed in Fig. 2.
Fitting was performed by using two Lorentzian curves convolved with a Gaussian peak function (FWHM $\sim$ 13.5 $\mu$eV) that represents the spectrometer response function.
For the emitter spectra ($\alpha =50.5^\circ$), if the cavity peak is not found, we reduce the number of Lorentzian peak to one. 
The results (integrated peak intensities, peak energies and linewidths) plotted as a function of cavity detunings are summarized in Fig.~\ref{fig:fit_f2}.
Through the fit, we can deduce some of cavity QED parameters.
$|g|$ = 41$\mu$eV is determined from the value of vacuum Rabi splitting 
and $\kappa$ = 66$\mu$eV is obtained from the fitting linewidth for the cavity-like polariton branch under a large detuning condition (see the linewidth plot for the cavity data ($\alpha =50.5^\circ$)）.
For the peak positions and linewidths, significant differences between the two sets of data are not found in their values at each detuning. 
However, the two intensity curves a show clear difference: the total intensities of emitter spectra become much weaker as approaching to the resonance. 
This can also be confirmed by the colorplot in Fig. 2. 

\subsection*{V. Theoretical modeling}

In this subsection, we derive a set of equations used for calculating the spectra in the main text. 
We consider a cavity QED system composed of a cavity mode (at frequency of $\omega _{c}$) and a two level emitter ($\omega _{a}$).
The system Hamiltonian of this model can be described as follows~\cite{Mandel1995},
\begin{eqnarray} 
H_S &=& H_0 + H_I \\
H_0 &=& \hbar \omega _{c} a^{\dagger}_{c}a_{c} + \hbar \omega _{a} \sigma ^{\dagger} \sigma \\
H_I &=&  - [\sigma \vector{d_{a}} + \sigma ^{\dagger} \vector{d_{a}}^{*}] \cdot \vector{E_{c}}.
\label{eq:Hjc}
\end{eqnarray}
Here, $a_{c}$ is the annihilation operator of the cavity photon, $\sigma$ the lowering operator for the two level system, \vector{d_{a}} the atomic dipole moment and
\vector{E_{c}} local cavity field vector at the location of the atom.
We quantize the cavity mode to be $\vector{E_{c}} = i\vector{\epsilon_{c}}(a_{c} - a_{c}^{\dagger})$, 
where $\vector{\epsilon_{c}}$ is a vector parallel to the cavity field local polarization at the atom location and 
has a maximum length, $\sqrt{\frac{\hbar \omega _{c}}{2\epsilon \epsilon _0 V}}$, if the atom is positioned in the field maximum. 
Here $V$ is the cavity mode volume and $\epsilon$ and $\epsilon _0$ are the relative and vacuum permittivity, respectively. 
Using Eq.~\ref{eq:dipole} and rotating wave approximation, the interaction part of the above Hamiltonian becomes,   
\begin{equation}
H_I = i \hbar \tilde{g} (\cos \beta \cos \theta _{a} + \sin \beta \sin \theta _{a} e^{\pm  i \phi_{QD}}) a_{c}^{\dagger} \sigma +h.c.,
\label{eq:Hi}
\end{equation}
where $\tilde{g} = |\vector{d_{a}}| |\vector{\epsilon_{c}}| /\hbar $, and $\beta$ is the angle between local field vector $\vector{\epsilon_{c}}$ and the x axis. 
The vacuum Rabi splitting induced by this Hamiltonian is $2g = 2\tilde{g} |\cos \beta \cos \theta _{a} + \sin \beta \sin \theta _{a} e^{\pm  i \phi_{QD}}|$. 
From the fits to the experimental VRS data, we obtaine $g$ = 41 $\mu$eV. 
The values of $\theta _{a}$ and $\phi_{QD}$ are also experimentally obtained by fitting to Fig.~\ref{fig:qd}(b). 
Using these values, we obtain possible values of $\beta$ = -87$^\circ$, -31$^\circ$, 93$^\circ$, 149$^\circ$. 
For the reproduction of experimental data, $\beta$ needs to be -31$^\circ$. 
$\beta$ can also be set to an equivalent value of 149$^\circ$ (=-31$^\circ$+180$^\circ$), which corresponds to the case $\vector{\epsilon_{c}} \to -\vector{\epsilon_{c}}$,
which does not make any difference to the reproduced results. 

Next, we consider the system-free space vacuum field interaction by introducing the following Hamiltonians,
\begin{eqnarray} 
H_{R} &=& \hbar \sum^{}_{\vector{k}, \lambda} \omega _{\vector{k}, \lambda} b^{\dagger}_{\vector{k}, \lambda}b_{\vector{k}, \lambda}, \\
H_{SR} &=& \hbar \sigma \sum^{}_{\vector{k}, \lambda} \tilde{\gamma}_{\vector{k}, \lambda}^{*} b^{\dagger}_{\vector{k}, \lambda} + \hbar a_{c} \sum^{}_{\vector{k}, \lambda} \tilde{\kappa}_{\vector{k}, \lambda}^{*} b^{\dagger}_{\vector{k}, \lambda}  +h.c.,
\label{eq:Hsr}
\end{eqnarray}
where $H_R$ and $H_{SR}$ respectively correspond to the free space reservoir energy and the system-reservoir interaction energy. 
The latter Hamiltonian is formed under the rotating wave approximation. 
$b_{\vector{k}, \lambda}^{\dagger}$ is the creation operator for the free space field that is specified with a wave vector of $\vector{k}$ and associated polarization $\lambda$ and has a frequency of $\omega _{\vector{k}, \lambda}$. 
$\tilde{\gamma}_{\vector{k}, \lambda}$ and $\tilde{\kappa}_{\vector{k}, \lambda}$ are the coupling constants between the free space field and the cavity photon and two level system, respectively.
Here, the free space reservoir includes the modes in the adjacent waveguide. Coupling between the system and the waveguide is assigned to corresponding wavevectors and polarizations.
We quantize the free space mode as,
\begin{equation}
\vector{E_{f}} = i\sum^{}_{\vector{k}, \lambda} \vector{e} _{\vector{k}, \lambda} \epsilon ^{f} _{\vector{k}, \lambda} (b_{\vector{k}, \lambda} e^{i\vector{k} \cdot \vector{r}} - b^{\dagger}_{\vector{k}, \lambda} e^{-i\vector{k} \cdot \vector{r}}) ,
\label{eq:Ef0}
\end{equation}
where $\epsilon ^{f} _{\vector{k}, \lambda}$ is the single photon field amplitude for a given quantization volume $V_Q$ and can be taken as real, namely $\epsilon ^{f} _{\vector{k}, \lambda}=\sqrt{\frac{\hbar \omega _{\boldsymbol{k}, \lambda}}{2 \epsilon _0 V_Q}}$.
$\vector{e} _{\vector{k}, \lambda} $ is a unit polarization vector orthogonal to ${\vector{k}}$ and $e^{i\vector{k} \cdot \vector{r}}$ expresses the spatial distribution.
Since the emitter and the free space couple through the dipole interaction, $-[\sigma \vector{d_a} + \sigma ^{\dagger} \vector{d_a}^*]\cdot \vector{E_f}$, 
the emitter-free space dipole coupling constant is expressed as,
\begin{eqnarray}
\hbar \tilde{\gamma}_{\vector{k}, \lambda} &=& -i e^{i\vector{k} \cdot \vector{r_a}} \epsilon ^{f} _{\vector{k}, \lambda}  \vector{e} _{\vector{k}, \lambda} \cdot \vector{d_{a}}^*, \label{eq:gamma1} 
\end{eqnarray}
where $\vector{r_a}$ is the atom location.
Meanwhile, the form of cavity-free space interaction, in other words the expression for $\tilde{\kappa}_{\vector{k}, \lambda}$, may be nontrivial~\cite{Viviescas2003}. 
For the sake of later discussion, we also separate $\vector{E_{f}}$ into positive and negative frequency parts,
\begin{eqnarray}
\vector{E_{f}} &=& \vector{E^+_{f}} + \vector{E^-_{f}}, \nonumber \\
\vector{E^+_{f}} &=& i\sum^{}_{\vector{k}, \lambda} \vector{e} _{\vector{k}, \lambda}  \epsilon ^{f} _{\vector{k}, \lambda} b_{\vector{k}, \lambda} e^{i\vector{k} \cdot \vector{r}},  \label{eq:Ef} \\
\vector{E^-_{f}} &=& -i\sum^{}_{\vector{k}, \lambda} \vector{e} _{\vector{k}, \lambda}  \epsilon ^{f} _{\vector{k}, \lambda} b^{\dagger}_{\vector{k}, \lambda} e^{-i\vector{k} \cdot \vector{r}}. \nonumber
\end{eqnarray}
Following a standard procedure with the Born and Markov approximations, we obtain a quantum master equation that describes the time evolution of the system density operator, $\rho$, 
\begin{eqnarray} 
\frac{d \rho}{d t} &=& -\frac{i}{\hbar}[H_{S},\rho ] + L\rho +L'\rho, \label{eq:master} \\
L\rho &=& \frac{\gamma}{2}(2\sigma \rho \sigma^{\dagger} - \sigma^{\dagger} \sigma \rho -\rho \sigma^{\dagger} \sigma ) \\
		 & &+ \frac{\kappa}{2}(2a_c\rho a_c^{\dagger} - a_c^{\dagger} a_c \rho -\rho a_c^{\dagger} a_c), \nonumber 
\end{eqnarray}
with
\begin{eqnarray} 
\gamma &=&  \sum^{}_{\vector{k}, \lambda} \pi |\tilde{\gamma}_{\vector{k}, \lambda}|^2 \delta ( \omega _{\vector{k}, \lambda} -\omega _a), \label{eq:decay1} \\ 
\kappa &=& \sum^{}_{\vector{k}, \lambda} \pi |\tilde{\kappa}_{\vector{k}, \lambda}|^2 \delta ( \omega _{\vector{k}, \lambda} -\omega _c). \label{eq:decay2}
\end{eqnarray}
These values should match with the experimentally measured spontaneous emission rate ($\gamma$ = 0.28 $\mu$eV) and cavity decay rate ($\kappa$ = 66 $\mu$eV). 
The Lamb shifts of the cavity and emitter are assumed to be absorbed in their original frequencies. 
We also add other Liouvillian terms:
\begin{eqnarray} 
L'\rho &=& \frac{P_a}{2}(2\sigma ^\dagger \rho \sigma  - \sigma \sigma ^\dagger \rho -\rho \sigma \sigma^\dagger) \nonumber \\
		  & &+ \gamma_{ph}(2\sigma ^\dagger \sigma \rho \sigma ^\dagger \sigma - \sigma ^\dagger \sigma \rho -\rho \sigma ^\dagger \sigma ). \nonumber 		  
\label{eq:master_add}
\end{eqnarray}
These are for introducing an incoherent atom pump ($P_a$) and an atomic phase decay ($\gamma_{ph}$).
For all the simulations, we set $P_a = $0.065$ \mu eV$ and $\gamma_{ph} = $3$ \mu eV$.

Now we consider the state of field released from the system into the free space through the system-reservoir interaction~\cite{Yamaguchi2010,Scully1997}. 
The Heisenberg equation of motion of the free space field operator is,
\begin{eqnarray} 
\frac{d}{dt}b_{\vector{k}, \lambda}(t) &=& \frac{i}{\hbar}[H_S+H_R+H_{SR},b_{\vector{k}, \lambda}] \\ 
&=& -i( \omega _{\vector{k}, \lambda} b_{\vector{k}, \lambda}(t) + \tilde{\gamma}_{\vector{k}, \lambda}^* \sigma (t) + \tilde{\kappa}_{\vector{k}, \lambda}^* a_c (t) ). \nonumber
\label{eq:heisenberg}
\end{eqnarray}
We can formally integrate this differential equation to obtain an expression for $b_{\vector{k}, \lambda}(t)$,
\begin{eqnarray} 
b_{\boldsymbol{k}, \lambda}(t) &=& b_{\boldsymbol{k}, \lambda}(0) e^{-i \omega _{\boldsymbol{k}, \lambda} t} \nonumber \\ &-& i \tilde{\gamma}_{\boldsymbol{k}, \lambda}^* \int _0^t dt' \sigma (t') e^{-i \omega _{\boldsymbol{k}, \lambda} (t-t')} \\
&-& i \tilde{\kappa}_{\boldsymbol{k}, \lambda}^* \int _0^t dt' a_c (t') e^{-i \omega _{\boldsymbol{k}, \lambda} (t-t')} \nonumber
\label{eq:b}
\end{eqnarray}
The first term corresponds to thermal photons initially existing in the free space reservoir and can be neglected. 
The second and third terms respectively correspond to the free space fields excited by spontaneous decay of the emitter and cavity.
Plugging this back to Eq. ~\ref{eq:Ef}, we obtain the expression for the positive part of free space electric field, $\vector{E^+_{f}}(\vector{r},t)$ at a position $\vector{r}$ :
\begin{eqnarray} 
& &\vector{E^+_{f}}(\vector{r},t) = \vector{E^{a+}_{f}}(\vector{r},t) + \vector{E^{c+}_{f}}(\vector{r},t), \\
& &\vector{E^{a+}_{f}} = \sum^{}_{\vector{k}, \lambda} \vector{e} _{\vector{k}, \lambda} \epsilon ^{f} _{\vector{k}, \lambda} \tilde{\gamma}_{\vector{k}, \lambda}^* \int_0^t dt' \sigma (t') e^{-i( \omega _{\boldsymbol{k}, \lambda} (t-t') -\boldsymbol{k} \cdot \boldsymbol{r}) }, \nonumber \\
& &\vector{E^{c+}_{f}} = \sum^{}_{\vector{k}, \lambda} \vector{e} _{\vector{k}, \lambda}  \epsilon ^{f} _{\vector{k}, \lambda} \tilde{\kappa}_{\vector{k}, \lambda}^* \int_0^t dt' a_c (t') e^{-i( \omega _{\boldsymbol{k}, \lambda} (t-t') -\boldsymbol{k} \cdot \boldsymbol{r}) }, \nonumber
\label{eq:Ef2}
\end{eqnarray}
where we can separately discuss the atomic ($\vector{E^{a+}_{f}}(\vector{r},t)$) and cavity ($\vector{E^{c+}_{f}}(\vector{r},t)$) fields. 

We proceed with the theory for the atomic contribution first. 
We take the atomic position as origin ($\vector{r_a} = \vector{0} $) without loss of generality and use a polar coordinate that is related to the Cartesian coordinate by $\vector{k}=k(\sin\theta\cos\phi,\sin\theta\sin\phi,\cos\phi)$. 
We only consider the farfield detected at a $\vector{r}$ that is parallel to $z$-axis ( $\vector{k} \cdot \vector{r} = kr\cos \theta $). At $\vector{r}$, the polarization of the farfield will be parallel to $\{ \hat{x},\hat{y} \}$.
Moreover, the single photon field amplitude can be approximated to be $\epsilon ^{f} _{\vector{k}, \lambda} \sim \epsilon ^{f} $. 
This is valid as far as the energy of the free space field spreads closely around $\omega _a$ for any $\{ \vector{k}$, $\lambda \}$ mode that contributes to $\vector{E^+_{f}}$.
The summation over $\vector{k}$ can be replaced by $\frac{V_Q}{(2\pi)^3}\int dk^3$ and $dk^{3}=k^{2}\sin\theta dkd\theta d\phi$. 
Now, the atomic decay contribution reads
\begin{eqnarray}
\vector{E^{a+}_{f}} &=& \epsilon ^{f} \sum^{}_{\lambda} \int  \tilde{\gamma}_{\vector{k},\lambda}^* \vector{e} _{\vector{k}, \lambda}  k^2\sin\theta dk d\theta d\phi \nonumber \\ 
&\times & \int_0^t dt' \sigma (t') e^{-i \omega _{\boldsymbol{k}, \lambda} (t-t') + ikr\cos \theta } .
\label{eq:Efa1}
\end{eqnarray}
Following a conventional process~\cite{Yamaguchi2010,Scully1997} and by using Eq. ~\ref{eq:gamma1} and the Weisskopf-Wigner approximation, we obtain the expression for the farfield ($(kr)^{-1}<<1$):
\begin{eqnarray}
& & \vector{E^{a+}_{f}} = \frac{1}{8 \pi \epsilon_{0} r} \frac{\omega_{a}^{2}}{c^{2}} \sigma (t-\frac{r}{c}) \vector{d_a} \label{eq:Efa} \\
&=& -i \sum _{\lambda=x,y} \frac{1}{8} \vector{e}_{\lambda} \tilde{\gamma} _{\lambda}^{*} \frac{\frac{2\pi c}{\omega_{a}}}{r} \epsilon_{f} D(\omega _a) \sigma(t-\frac{r}{c}), \label{eq:Efa2}
\end{eqnarray}
where $c$ is the speed of light and
\begin{eqnarray}
& & \hbar \tilde{\gamma} _{x} = i \epsilon _f |\vector{d_a}| \cos \theta _a ,\\
& & \hbar \tilde{\gamma} _{y} = i \epsilon _f |\vector{d_a}| e^{\pm i \phi_{QD}} \sin \theta _a .
\end{eqnarray}
Eq.~\ref{eq:Efa2} has a simple form: coupling constant $\times$ wavelength ($2 \pi c/\omega_a$) over distance ($r$) $\times$ single photon field amplitude $\times$ density of states ($D(\omega _a)=\frac{V_{Q}\omega_{a}^{2}}{\pi^{2}c^{3}}$) $\times$ system operator. 

Second, we apply a similar deformation for the cavity leakage contribution and obtain, 
\begin{eqnarray}
\vector{E^{c+}_{f}} &=& \epsilon ^{f} \sum^{}_{\lambda} \int  \tilde{\kappa}_{\vector{k},\lambda}^* \vector{e} _{\vector{k}, \lambda}  k^2\sin\theta dk d\theta d\phi \nonumber \\ 
&\times & \int_0^t dt' a_c (t') e^{-i \omega _{\boldsymbol{k}, \lambda} (t-t') + ikr\cos \theta } .
\label{eq:Efc1}
\end{eqnarray}
In order to be in line with the experiment, we neglect the $y$-polarization contribution in the farfield and assume that the cavity mode only couples to the $x$-polarized free space field, $\tilde{\kappa}_{\vector{k},\lambda}=\tilde{\kappa}_{\vector{k},x}$.
Moreover, we assume that $\tilde{\kappa}_{\vector{k},\lambda}$ has no dependence on $\vector{r}$ and takes the form, $\tilde{\kappa}_{\vector{k},x} = \bar{\kappa} f(\theta) g(\phi)$. 
This is for including the effect of inhomogeneous radiation pattern of the cavity mode.
If $f(\theta =0) = o(1), \int_0^{2\pi}g(\phi)d\phi = o(1)$ and $\int_0^{\pi} \frac{d f(\theta )}{d \theta} e^{ikr\cos \theta}d\theta \sim o(\frac{1}{kr})$, then $\vector{E^{c+}_{f}}$ in the farfield region becomes
\begin{equation}
\vector{E^{c+}_{f}} = -i \frac{1}{8} \hat{x} \Lambda _c \bar{\kappa}^{*} \frac{\frac{2\pi c}{\omega_{c}}}{r} \epsilon_{f} D(\omega _c) a(t-\frac{r}{c}),
\end{equation}
where $\Lambda _c$  represents the radiation pattern effect and $\Lambda _c= \frac{f(0)}{2\pi}\int_0^{2\pi}g(\phi)d\phi$. 
$\bar{\kappa}$ is the cavity-free space coupling constant when assuming homogeneous coupling over any $\vector{k}$.

Now we separate the $x$ and $y$ polarized contribution in the farfield, namely, $\vector{E^{a+}_{f}}(\vector{r},t) = \vector{E^{a+}_{x}} + \vector{E^{a+}_{y}}$ and  $\vector{E^{c+}_{f}}(\vector{r},t) = \vector{E^{c+}_{x}}$ and
slightly modify the expressions for each component vector as,
\begin{eqnarray}
\vector{E^{a+}_{x}} &=& \frac{c \epsilon_{f} \sqrt{\pi D(\omega _a)}}{4\hbar \omega_{a} r} \sqrt{\gamma} \cos \theta _a \hat{x} \sigma (t), \label{eq:Efax0} \\
\vector{E^{a+}_{y}} &=& \frac{c \epsilon_{f} \sqrt{\pi D(\omega _a)}}{4\hbar \omega_{a} r} \sqrt{\gamma} \sin \theta _a e^{i\pm \phi_{QD}}\hat{y} \sigma (t), \label{eq:Efay0} \\
\vector{E^{c+}_{x}} &=& \frac{c \epsilon_{f} |\Lambda _c| \sqrt{\pi D(\omega _c)}}{4\hbar \omega_{c} r} (-i \sqrt{\kappa} e^{-i\theta _c}) \hat{x} a_c(t), \label{eq:Efcx0} 
\end{eqnarray}
where we redefined the time by setting $t-\frac{r}{c} \rightarrow t$ and used the fact that $\epsilon_f |\vector{d_a}|/\hbar = \sqrt{\frac{\gamma}{\pi D(\omega_a)}}$ and $|\bar{\kappa}| = \sqrt{\frac{\kappa}{\pi D(\omega_c)}}$.
We also introduced the cavity-free space coupling phase, $\theta _c$, by setting $\bar{\kappa} \Lambda _c^*=|\bar{\kappa}||\Lambda _c|e^{i\theta _c}$. 
In experiments, we measured the field just around a position $\vector{r} \parallel \hat{z}$ and used a very small numerical aperture ($\sim 0.1$), in which the farfield can be assumed to marginally change.
Then, the measured field, $\vector{E^{+}_{D}}$, will be expressed as a sum of $\vector{E^{a+}_{f}} $ and $\vector{E^{c+}_{f}} $, after multiplying the detection area.
We further introduce factors $A'$ and $B'$ in order to take into account the difference of the amount of field passing through the aperture:
\begin{equation} 
\vector{E^{+}_{D}} = A' \vector{E^{c+}_{f}} + B' \vector{E^{a+}_{f}},
\end{equation}
and each polarization component reduces to
\begin{eqnarray}
{{E}^{a+}_{x}} &=& \sqrt{A} \sqrt{\gamma} \cos \theta _a \sigma(t), \label{eq:Efax} \\
{{E}^{a+}_{y}} &=& \sqrt{A} \sqrt{\gamma} \sin \theta _a e^{i\pm \phi_{QD}} \sigma(t), \label{eq:Efay} \\
{{E}^{c+}_{x}} &=& \sqrt{B} (-i \sqrt{\kappa} e^{-i\theta _c}) a_c(t), \label{eq:Efcx} 
\end{eqnarray}
where $A$ and $B$ are parameters that correspond to photon energy observable within the detection area in the unit time due to one photon leakage, respectively for the atomic and cavity contributions.
The ratio $A/B$ (=2.85) can be determined from fitting to experimental data (Fig. 3(c)).
Referring to Fig. 4, the fields after passing through the HWP and PBS reads 
\begin{equation} 
{{E}^{+}_{P,D}} = {{E}^{c+}_{x}}\cos \Theta + {{E}^{a+}_{x}} \cos \Theta + {{E}^{a+}_{y}} \sin \Theta .
\end{equation}
$\Theta$ is the angle between the $x$ axis and the rotated $\vector{E^{+}_{x}}$ field. 
Then, the detected field intensity, ${E}^{+}_{P,D}{E}^{-}_{P,D}$, can be expressed as
\begin{eqnarray}
{E}^{+}_{P,D}{E}^{-}_{P,D} &=& ({E}^{c+}_{x} \cos \Theta)\times h.c. \\
&+& ({E}^{a+}_{x}\cos \Theta +{E}^{a+}_{y}\sin \Theta )\times h.c. \\
&+& ({E}^{c+}_{x}{E}^{a-}_{x} +{E}^{c-}_{x}{E}^{a+}_{x}) \cos ^2 \Theta \\
&+& ({E}^{c+}_{x}{E}^{a-}_{y} +{E}^{c-}_{x}{E}^{a+}_{y}) \sin \Theta \cos \Theta 
\end{eqnarray}
The first and second term respectively correspond to the pure cavity and atomic radiation. 
The third and fourth term are from the interference between $x$-polarized cavity and emitter, and between $x$-polarized cavity and y-polarized emitter, respectively.
Without the insertion of the polarizer, the latter term does not exist, so that we can call it polarizer-induced interference term. 
According to the Wiener-Khinchin theorem, emission spectra can be calculated by the Fourier transform of the two time correlation function, $\langle {{E}^{+}_{P,D}}(t) {{E}^{-}_{P,D}}(t+\tau)\rangle$. 
Therefore, the spectrum in the steady state ($t\rightarrow\infty$), $S(\omega)$, reads
\begin{eqnarray}
S(\omega)&=&\lim_{t \to \infty}\frac{1}{2\pi} \int d\tau e^{i\omega \tau} \langle \tilde{E}^{+}_{P,D}(t) \tilde{E}^{-}_{P,D}(t+\tau) \rangle \\
&=&S_c(\omega)+S_a(\omega)+S_{I1}(\omega)+S_{I2}(\omega),
\end{eqnarray}
where
\begin{widetext}
\begin{eqnarray}
S_c(\omega)&=&\frac{B\kappa\cos^2\Theta}{2\pi} \int d\tau e^{i\omega \tau} \langle a^\dagger_c (t) a_c(t+\tau) \rangle , \label{eq:Sc} \\
S_a(\omega)&=&\frac{A\gamma |\cos\theta_a\cos\Theta+e^{i\phi_{QD}} \sin\theta_a\sin\Theta |^2}{2\pi} \int d\tau e^{i\omega \tau} \langle \sigma^\dagger (t) \sigma (t+\tau) \rangle ,\label{eq:Sa} \\
S_{I1}(\omega)&=&\frac{\sqrt{AB}\sqrt{\kappa\gamma p} \cos\theta_a\cos^2\Theta}{2\pi} \int d\tau e^{i\omega \tau} \{  e^{-i(\frac{\pi}{2}+\theta_c)}\langle \sigma^\dagger (t) a_c(t+\tau) \rangle + e^{i(\frac{\pi}{2}+\theta_c)}\langle a_c^\dagger (t) \sigma(t+\tau) \rangle  \} , \label{eq:SI1} \\
S_{I2}(\omega)&=&\frac{\sqrt{AB}\sqrt{\kappa\gamma p} \sin\theta_a\sin\Theta \cos\Theta}{2\pi} \int d\tau e^{i\omega \tau} \{  e^{-i(\frac{\pi}{2}+\theta_c \pm \phi_{QD})}\langle \sigma^\dagger (t) a_c(t+\tau) \rangle + e^{i(\frac{\pi}{2}+\theta_c \pm \phi_{QD}))}\langle a_c^\dagger (t) \sigma(t+\tau) \rangle  \}. \nonumber \\
\label{eq:SI2}
\end{eqnarray}
\end{widetext}
$p$ is introduced to account for the field intensity distribution overlap between the cavity and emitter contributions within the aperture ($p=0$, no overlap, $p=1$, perfect overlap). 

We solved the master equation (Eq.~\ref{eq:master}) for the steady state by setting $\frac{d}{dt}=0$, and then calculated the two time correlation functions using the quantum regression theorem.
By Fourier transforming the two time correlation functions, as in Eq.\ref{eq:Sc}-\ref{eq:SI2}, we obtain the spectra after convolving with the detector response function.   
The spectra is calculated as the average of the results for the two signs of $\phi_{QD}$, because our QD can be assumed to randomly choose either sign during the measurements.

Since we treat the interference, the phase of coupling constants between the emitter, cavity and free space, namely the arguments of $g$, $\tilde{\kappa}$ and $\tilde{\gamma}$, significantly affect on the result.  
The emitter couples with the cavity and the free space through the well-known dipole interaction that naturally incorporates the phase information in the coupling constants, $g$ and $\tilde{\gamma}$.
However, the cavity-free space coupling constant, $\tilde{\kappa}$, in general has a nontrivial form~\cite{Viviescas2003} and is difficult to determine its phase, which is, then, often left ambiguous in the literature. 
Our work here accesses this issue and can experimentally evaluate the phase of cavity-free space coupling, namely $\theta_c$. 
In this discussion, we can neglect the difference in the propagation phase between the two detection channels (quantified by the distance from the radiation source to the detector), 
since the QD located near the cavity center and both of them radiate from a well localized region ($\ll$ 1 $ \mu$m) that is much smaller than the emission wavelength ($\sim$ 1 $\mu$m). 
When reproducing the measured spectra, we had to take $|\theta_c| < 30^\circ$ for $\beta =-31^\circ $ and the best simulation was obtained when $\theta_c \sim 0^\circ$ 
($\theta_c \sim 180^\circ$ for $\beta =149^\circ $).
Therefore, we took $\theta_c = 0^\circ$ for simulating all the spectra in this work. 
According to J. T. Shen and S. Fan~\cite{Shen2009}, $\theta_c = 0^\circ$ indicates that the cavity holds time reversal symmetry and a mirror symmetry in the direction of photon leakage. 
It is worth noting that the value largely deviates from the coupling phase estimated by numerically solving Maxwell's equation ($\sim$ -85$^\circ$) for our cavity structure, as plotted in Fig.~\ref{fig:FF}(e). 
Further detailed studies will be necessary to verify the actual coupling phase between the cavity and free space.
It will also be interesting to verify the coupling phase in other radiation directions not studied in this work ($\parallel z$).

\subsection*{VI. Emitter-cavity complementarity in VRS spectra}

When under weak pumping, the emitter and cavity photon are in the linear strong coupling regime of cavity QED. 
In this regime, an energy quantum injected into either the emitter or cavity will exhibit the same fundamental vacuum Rabi oscillation in the time domain.
Therefore, the VRS spectrum of the emitter under weak emitter-driven conditions should coincide with the VRS spectrum of the cavity under weak cavity-driven conditions. 
In other words, the emitter and cavity photon hold a complementarity~\cite{Berman1994}.
The main difference induced by the different pumping conditions is a different initial condition of the vacuum Rabi oscillation, 
which is thus assumed to be the dominant cause of the difference between the emitter and cavity VRS spectra under the emitter-driven conditions, as observed in the main text. 

\begin{figure}
\centering
\includegraphics[width=\linewidth ]{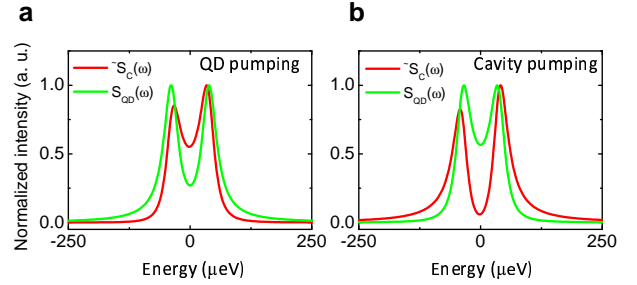}
\caption{{\bf Comparisons of VRS spectra under two different pump conditions.} 
{\bf a}, Normalized VRS spectra under emitter-driven conditions, as obtained by the numerical simulations (the same plots in Fig.3(a) and (b) in the main text). 
{\bf b}, Normalized VRS spectra under cavity-driven conditions. 
The spectra were simulated with exactly the same procedure as in {\bf a}, albeit with only the cavity being pumped. 
In this simulation, we set the emitter pumping rate to zero ($P_a = 0$) and introduced an incoherent cavity pumping by coupling a finite temperature reservoir to the cavity mode.
}
\label{fig:compli}
\end{figure}

In order to verify this, we performed numerical simulations of VRS spectra for two different pumping conditions. 
For the emitter-driven condition, we used exactly the same pumping condition as in Fig.3(a) and (b) in the main text, and the results are replotted in Fig.~\ref{fig:compli}(a).
For the cavity-driven condition, we switched off the emitter pumping ($P_a = 0$) and introduced an incoherent cavity pumping ($P_c$) through the following Liouvillian, which was added to the master equation (Eq.~\ref{eq:master}):
\begin{eqnarray} 
L''\rho &=& \frac{P_c}{2}(2a_c\rho a_c^{\dagger} - a_c^{\dagger} a_c \rho -\rho a_c^{\dagger} a_c) \\
 		  & &+ \frac{P_c}{2}(2a_c^{\dagger} \rho a_c -  a_c a_c^{\dagger} \rho -\rho  a_c a_c^{\dagger}) , \nonumber 
\label{eq:cavpump}
\end{eqnarray}
where we set $P_c =$0.065$\mu eV$. 
The simulated spectra for the two detection channels are plotted in Fig.~\ref{fig:compli}(b).
As expected, the emitter VRS spectrum exhibits a narrower splitting and a shallower central dip when the cavity is pumped. 
The remaining asymmetry in the mainly-cavity channel spectrum is due to the effect of the interference described in the main text.

\bibliography{QDC_arXiv}
\bibliographystyle{apsrev4-1}



\end{document}